\documentstyle[emulateapj]{article}

\articleid{11}{14}
 
\def\lsim{\mathrel{\rlap{\lower 4pt \hbox{\hskip 1pt $\sim$}}\raise 1pt \hbox
        {$<$}}}
\def\gsim{\mathrel{\rlap{\lower 4pt \hbox{\hskip 1pt $\sim$}}\raise 1pt \hbox
        {$>$}}}

\lefthead{Maeda et al.}
\righthead{Nucleosynthesis and Spectra of SN1998bw}

\newcommand{\etal}{et~al.\ }

\newcommand{\OI}{O~{\sc i}}
\newcommand{\MgI}{Mg~{\sc i}}
\newcommand{\CaII}{Ca~{\sc ii}}
\newcommand{\CoII}{Co~{\sc ii}}
\newcommand{\FeI}{Fe~{\sc i}}
\newcommand{\FeII}{Fe~{\sc ii}}
\newcommand{\FeIII}{Fe~{\sc iii}}

\begin{document}

\submitted{To appear in the January 2002 issue of ApJ (Vol. 565)}

\title{Explosive Nucleosynthesis in Aspherical Hypernova Explosions \\and 
Late Time Spectra of SN1998bw}

\author{Keiichi Maeda\altaffilmark{1}, Takayoshi Nakamura\altaffilmark{1}, 
Ken'ichi Nomoto\altaffilmark{1,5}, Paolo A. Mazzali\altaffilmark{2,5}, 
Ferdinando Patat\altaffilmark{3}, Izumi Hachisu\altaffilmark{4}}

\altaffiltext{1}{Department of Astronomy, School of Science, 
University of Tokyo, Bunkyo-ku, Tokyo 113-0033, Japan}
\altaffiltext{2}{Osservatorio Astronomico di Trieste, 
via G. B. Tiepolo 11, I-34131 Trieste, Italy}
\altaffiltext{3}{European Southern Observatory, K. Schwarzschild Str. 
2, D-85748 Garching, Germany}
\altaffiltext{4}{Department of Earth Science and Astronomy, 
College of Arts and Science, University of Tokyo, Meguro-ku, 
Tokyo 153-0041, Japan}
\altaffiltext{5}{Research Center for the Early Universe,
School of Science, University of Tokyo, Bunkyo-ku, Tokyo 113-0033, Japan \\
emails: maeda@astron.s.u-tokyo.ac.jp, 
nakamura@astron.s.u-tokyo.ac.jp, 
nomoto@astron.s.u-tokyo.ac.jp, 
mazzali@ts.astro.it, 
fpatat@eso.org, 
hachisu@chianti.c.u-tokyo.ac.jp}

\begin{abstract}
Aspherical explosion models for the hypernova 
(hyper-energetic supernova) SN~1998bw are presented. 
Nucleosynthesis in aspherical explosions is examined with a two-dimensional 
hydrodynamical code and a detailed nuclear reaction network. 
Aspherical explosions lead to a strong $\alpha$-rich freezeout, thus 
enhancing the abundance ratios 
[$^{44}$Ca, $^{48}$Ti, and $^{64}$Zn / Fe] in the ejecta. 
The nebular line profiles of the Fe-dominated blend near 5200 \AA\ and of 
[\OI] 6300,6363 \AA\ are calculated and compared with the observed 
late time spectra of SN 1998bw. 
Compared with the spherical model, the unusual features of the 
observed nebular spectra can be better explained if SN 1998bw 
is a strongly aspherical explosion with a kinetic energy of 
$\sim 10^{52}$ ergs viewed from near the jet direction. 
\end{abstract}


\section{Introduction}\label{sec:int}

The exceptionally bright Type Ic supernova (SN Ic) SN 1998bw was 
discovered as the probable optical counterpart of the gamma-ray burst 
GRB980425 (Galama et al. 1998). 
The early light curve and the spectra of SN 1998bw have been successfully 
modeled as the hyper-energetic explosion (kinetic energy 
$E \sim$ 4 $\times$ 10$^{52}$ ergs) of a massive C+O star 
(Iwamoto \etal 1998; Woosley \etal 1999; Branch 2001).  
In this paper the term "hypernova" is used to refer to a SN explosion with 
$E \gsim 10^{52}$ ergs, regardless of the nature of the central engine 
(Nomoto et al. 2001a). 

Despite the success of the hypernova model in reproducing the observed 
features of SN 1998bw at early times, some properties of the observed 
light curve and spectra at late times are difficult to explain.  
(1) The tail of the observed light curve declines more slowly than the 
synthetic curve, indicating that at advanced epochs $\gamma$-ray trapping is 
more efficient than expected (Nakamura \etal 2001a; Sollerman \etal 2000).  
(2) In the nebular epoch, the [\OI]6300\AA\ emission is narrower than the
emission near 5200\AA. As discussed in Mazzali \etal (2001), this 
latter feature is mostly due to a blend of [\FeII] lines. 
Mazzali \etal (2001) calculated synthetic nebular-phase spectra of SN~1998bw
using a spherically symmetric NLTE nebular code based on the deposition of 
$\gamma$-rays from $^{56}$Co decay in a nebula of uniform density and 
composition. 
They showed that the [\OI] and the [\FeII] features can only be reproduced 
if different velocities are assumed for the two elements. A significant amount 
of slowly-moving O is therefore necessary to explain the profile of the 
[\OI] line. 

Both these features are in conflict with what is expected from a spherically
symmetric explosion  model, where $\gamma$-ray deposition decreases with 
time and where iron is produced in the deepest layers and thus has a lower 
average velocity than oxygen. Mazzali \etal (2001) suggested that these are 
signatures of asymmetry in the ejecta.  Therefore in this paper we examine 
aspherical explosion models for hypernovae.

Aspherical explosions of massive stars have been investigated as possible 
sources of gamma-ray bursts (GRBs) (Woosley 1993; Paczynski 1998). 
MacFadyen \& Woosley (1999) showed numerically that the collapse of a 
rotating massive core can form a black hole with an accretion disk, 
while a jet emerges along the rotation axis. 
The jet produces a highly asymmetric explosion (Khokhlov \etal 1999).  
However, these studies did not calculate explosive nucleosynthesis, nor did 
they show spectroscopic and photometric features of aspherical explosions. 
Nagataki (2000) performed nucleosynthesis calculations for aspherical SN 
explosions to explain some features of SN~1987A, but he only addressed 
the case of a normal explosion energy. 

In the present study, we examine the effect of aspherical explosions
on nucleosynthesis in hypernovae.  We then investigate the degree of
asphericity in the ejecta of SN 1998bw, which is critically important
information to confirm the SN/GRB connection, by computing synthetic spectra
for the various models viewed with different orientations and comparing the
results with the observed late time spectra of SN~1998bw (Patat \etal 2001).

\section{Asymmetric Explosion Models}\label{sec:mod}

The first step of our calculation is the hydrodynamical simulation 
of the explosion with a 2D Eulerian hydrodynamical code based on 
Roe's scheme (Hachisu \etal 1992, 1994).  
Euler's equations are solved with a constant adiabatic index $\gamma$ = 4/3, 
which is a good approximation if the pressure is radiation-dominated. 
The effect of nuclear reactions on the hydrodynamics is negligible  
since the explosion energy is large.  
We use $120 \times 120$ meshes on a cylindrical $(r,z)$ coordinate system.  
The mesh size is linearly zoned and decreases inward, which gives a high 
resolution of the hydrodynamic evolution of the central regions where 
explosive nucleosynthesis takes place. 
We follow 190 test particles initially in the Si layer and 2250 particles 
initially in the C+O layer, tracking their density and temperature histories. 
These histories are then used to calculate the change in the chemical 
composition, using a reaction network including 222 isotopes up to 
$^{71}$Ge (Thielemann \etal 1996).

We construct several asymmetric explosion models for various 
combinations of the model parameters (Table 3). 
We use as progenitor the 16 $M_\odot$ He core of a 40 $M_\odot$ star 
(Nomoto \& Hashimoto 1988). 
This has a 13.8 $M_\odot$ C+O core, the same as that used in Iwamoto 
\etal (1998). 
We test three values of the final kinetic energy: 
$E = 1 \times 10^{51}$, $1 \times 10^{52}$, and $3 \times 10^{52}$ ergs. 
The hydrodynamical simulation is started by depositing the energy below 
the mass cut that divides the ejecta from the collapsing core.
The energy deposited is divided between thermal and kinetic energy, 
with various ratios. 
The asymmetry is generated by distributing the initial kinetic energy 
in an axisymmetric way. 
This is done by imposing different initial velocities in different 
directions: 
$v_z = \alpha \; z$ in the jet direction and 
$v_r = \beta \; r$ on the equatorial plane.  
The ratio $\alpha/\beta$ ranges from 16:1 to 1:1 (spherical case). 
The mass cut is set at $M_r = 2.4M_\odot$, so that the ejected mass of 
$^{56}$Ni is $\sim$ 0.4 $M_\odot$ to reproduce the peak of the light curve
(Nakamura \etal 2001a) in models A, B, C, E, and F.

\section{Nucleosynthesis}\label{sec:res}

Figure 1 and 2 show respectively the post-shock peak temperatures and
densities for the asymmetric hypernova model C in the direction of the jet and
perpendicular to it (with those for the spherically symmetric hypernova model
F also shown for comparison), and the isotopic composition of the ejecta of
model C.  In the $z$-direction, where the ejecta carry more kinetic energy, the
shock is stronger and post-shock temperatures are higher, so that explosive 
nucleosynthesis takes place in more extended, lower density regions compared 
with the $r$-direction.  Therefore, larger amounts of $\alpha$-rich freeze-out
elements, such as $^4$He and $^{56}$Ni (which decays into $^{56}$Fe via
$^{56}$Co) are produced in the $z$-direction than in the $r$-direction.  Also,
the expansion velocity of newly synthesized heavy elements is much higher in
the $z$-direction.  The velocity of elements ejected in the $z$-direction in
model C is  actually similar to the result of a spherical explosion with $E
\sim 3 \times 10^{52}$ ergs (Nakamura \etal 2001b), although the integrated
kinetic energy is only $E = 1 \times 10^{52}$ ergs.

In contrast, along the $r$-direction $^{56}$Ni is produced only in the
deepest layers, and elements ejected in this direction are mostly the
product of hydrostatic nuclear burning (O), with some explosive
oxygen-burning products (Si, S, etc). The expansion velocities are
much lower than in the $z$-direction.  

Figure 3 shows the 2D distribution of $^{56}$Ni and $^{16}$O in model C 
in the homologous expansion phase. 
Near the $z$-axis the shock is stronger and a low density, 
$^4$He-rich region is produced. 
$^{56}$Ni is distributed preferentially in this direction, but it is 
mostly located slightly off of it because the shock propagates 
laterally as it penetrates the stellar envelope. 
As a result, the distribution of heavy elements is elongated in the 
$z$-direction, while that of $^{16}$O is less aspherical.  
On the other hand, because the ejecta move more slowly in the $r$-direction, 
densities in this direction are higher than in the $z$-direction.

Tables 1 and 2 give respectively the detailed yields and the abundances 
of major stable isotopes relative to the solar values for model C 
(in Table 2, $\rm{[A/B]} \equiv \log_{10}\rm{(A/B)} - 
\log_{10}\rm{(A/B)}_\odot$, where A and B are nuclear mass fractions). 
The main characteristics can be summarized as follows 
(see also Nomoto et al. 2001b).  

(1) The complete Si-burning region is more extended for larger 
explosion energies. 
The aspherical explosion causes a region of higher entropy along the $z-$axis, 
which offers better conditions for the $\alpha$-rich freezeout (Fig. 1). 
The high entropy inhibits the production of $^{56}$Ni. 
Much $^4$He is left after the freezeout, so that the elements produced 
through $^{4}$He capture are very abundant in the deepest region along the 
$z$-axis (Fig. 2). 
This results in the enhancement of the elements synthesized in the deepest 
region, such as $^{44}$Ca (produced as $^{44}$Ti), $^{48}$Ti 
(as $^{48}$Cr), and elements heavier than $A \sim 58$. 
Because of the enhancement of these elements and the simultaneous suppression 
of $^{56}$Ni, the abundances of these elements relative to iron 
(e.g., [$^{44}$Ca, $^{48}$Ti, $^{64}$Zn /Fe]) are greatly enhanced. 
For more asymmetric explosion, the effect of $\alpha$-rich freezeout 
is even larger. 

(2) Incomplete Si-burning and O-burning regions are more extended for larger 
explosion energies (Nakamura et al. 2001b). 
This results in the enhancement of $^{28}$Si, $^{32}$S, $^{40}$Ca, $^{52}$Cr 
(produced as $^{52}$Fe), $^{54}$Fe, and in the reduction of O.  
Asphericity has little effect on the production of these elements. 

The most pronounced effect of asphericity is that elements produced by the 
strong $\alpha$-rich freezeout are greatly enhanced relative to iron 
(e.g., [Ti/Fe]). 
For other explosive burning products, the effect of a large explosion energy 
usually dominates over that of asphericity. 

\section{The Late Time Spectra of SN 1998bw}\label{sec:1998bw}

In order to verify the observable consequences of an axisymmetric explosion, 
we calculated the profiles of the [\FeII] blend and of [\OI] for models A-G. 
Line emissivities were obtained from a 1D NLTE nebular code 
(Mazzali et al. 2001), and the 
column densities of the various elements along different lines of sight were 
derived from the element distribution obtained from our 2D explosion models.  
Because we assume that the nebula is optically thin, the blended
nature of the emissions is automatically taken into account. 

These are compared to the 26 Nov 1998 spectrum of SN~1998bw.  We select this
spectrum because the wavelength of the 5200\AA\ feature, which was somewhat
redder at earlier epochs, at this and later epochs coincides with that of the
equivalent [\FeII] feature in the nebular spectra of SNe~Ia (Axelrod 1980),
indicating that other contributions (see Section 5) are now negligible.   
Table 3 gives the FWHM of our synthetic lines as a function of viewing angle.  
The corresponding observed value is 380\AA\ for the Fe-blend.
This is estimated assuming that the continuum level is the value around
5700\AA\ where the flux has the minimum value. 

The FWHM of the 5200\AA\ feature is narrower at this epoch than at earlier 
ones, and so is that of [\OI].  While for the former feature other
contributions may be responsible for a broader line at earlier epochs, in the
case of [\OI] the decreasing density of the outer envelope must be the
principal reason.  At late epochs the density of the outer envelope is expected
to be too small to trap the $\gamma$-rays.  Therefore we set the outermost
velocity of the emitting region to reproduce the FWHM of the [\OI] line
(150\AA), and then calculate the profiles of the [\FeII] blend.

The [\FeII] and [\OI] profiles for model C viewed at an angle of 15$^{\circ}$
from the jet direction and those for model F are compared to the observed 
spectrum of 26 Nov 1998 in Figure 4.  For model F in Fig. 4, the outermost
velocity of the emitting region is set to make the [\FeII] line as broad as
possible, because for this model we cannot get a reasonable fit for the [\OI]
line, which is always much broader than the observation.  Indeed, fitting the
[\OI] line was not possible for all models.  Among the hypernova models, in a
spherical explosion (model F) oxygen is located at higher velocities than iron,
and the [\OI] line is too broad for any choice of the outer velocity of the
emitting region.  This is because of the deficiency of oxygen with small
velocity along the line of sight.  Also, even though the Fe feature can be
wider than the O line if O and Fe are mixed extensively, the expected ratio of
the width of the Fe-blend and the O line even in a fully mixed model is $\sim
3:2$ (Mazzali \etal 2001).  This is the result of taking blending into
account, but giving all contributing lines the same intrinsic width. However,
the observed ratio is even larger, $\gsim 2:1$, implying that the [\FeII] lines
are intrinsically broader than the [\OI] ones.  Therefore the observed line
profiles are not explained with a spherical hypernova model.  The same is true
for the moderately asymmetric model E viewed near the equator.

In our aspherical explosion models Fe is distributed preferentially along 
the jet direction, and so a larger ratio of the Fe and O line widths can be
obtained. All the strongly aspherical hypernova models A, B and C, when viewed
from a near-jet direction, give line widths comparable to the observed values. 
The very energetic model B cannot reproduce the O line when viewed near the
equator, but this is because O is too fast near the equator and too depleted
near the poles to give a low-velocity component.

When the degree of asphericity is high and the explosion is viewed from near 
the jet direction, the component lines in the [\FeII] blend have double-peaked
profiles, the blue- and red-shifted peaks corresponding to matter situated 
in the two opposite lobes of the jet, where Fe is mostly produced.
Because of the high velocity of Fe, the peaks are widely separated, and the 
blend is wide (Fig. 4, model C).  In contrast, the [\OI] line is narrower and 
has a sharper peak, because O is produced mostly in the $r$-direction, 
at lower velocities and with a less aspherical distribution.

Figure 5 shows the [\OI] line for model C at different orientations. 
The mean expansion velocity of O is lower in the aspherical cases than 
in the spherical model, because in aspherical models low-velocity, 
high-density O-dominated matter is found near the center (Fig. 3). 
Therefore the width of the O line in the aspherical models (A, C) 
viewed from near the equator can also be comparable to the observed width. 
This, however, does not mean that the line profile is always sharp as seen in 
SN1998bw. As shown in Figure 5, when the angle is large, the line first 
broadens, and eventually it develops two peaks.  
The reason can be seen in Figure 3. The highest density region has a typical 
velocity $\sim$ 3000 km s$^{-1}$ along the $r$-axis in this model. 
This corresponds to a Doppler shift of $\sim$ 120\AA\ between the approaching 
and receding parts when the SN is viewed from the $r$-direction. 
Therefore, to produce the narrow and sharply peaked O line in a hypernova model,
the explosion must be aspherical and viewed from near the polar direction. 

Figure 6 shows the profiles of the [\FeII] blend viewed at 5$^{\circ}$ for 
various aspherical hypernova models (A, B, C and E). In all of these models, 
the computed O line reproduces the observed one as discussed above.  
The Fe-blends appear to reproduce the 5200\AA\ feature reasonably well. 
The profiles shows small peaks, which are not seen in the observations. 
These peaks are more pronounced in more asymmetric models.
Mixing of the ejecta may distribute the $^{56}$Ni to lower velocities, 
thus reducing the double-peaked profiles of the Fe lines.  
A spherical hypernova model (model F in Fig. 4) also gives a broad Fe line, 
and without the sharp peaks. However the O line is much too broad. 
Also, in this work we have used a spherical, uniform density nebular 
model to compute line emissivities. 
In an aspherical model, dense central and equatorial regions may have higher 
$\gamma$-ray trapping efficiency, which may result in stronger low-velocity 
line emission than in our model. 
Thus the component iron lines in the blend may have wider, flat-top profiles 
rather than double-peaked shapes, which could eliminate the minor peaks in 
the Fe blend seen in our present models. 
2D $\gamma$-ray trapping calculations are therefore needed to compute 
the detailed spectra and the light curve.

It is reasonable to think that asphericity reduces the energy below 
that estimated previously based on spherically symmetric models. 
To examine this, we turn now to the lower energy explosion models (D, G). 
The [\FeII] and [\OI] profiles for these models are shown in Figure 7. 
First of all, these models always give a narrower O line than the observed one. 
In these low energy models, in fact, the velocity of the ejecta is too small. 
The fastest-moving matter approaches the observer with a velocity of 3500 km 
s$^{-1}$ for model G, and 5000 km s$^{-1}$ in the case of the aspherical model 
(D) viewed from the $z$-direction. 
The observed FWHM (150\AA), however, indicates that there should be material 
moving faster than 7000 km s$^{-1}$. 
Also, because of the low velocity, the component [\FeII] lines are 
too narrow and do not blend to form a broad feature. 
From these arguments, 
we conclude that the explosion energy of SN1998bw should 
have been large, $E_{\rm K} \sim 1 \times 10^{52}$ergs. 

\section{Conclusion and Discussion}\label{sec:con}

We calculated the nucleosynthesis in aspherical hypernova explosions. 
We find that in such explosions Fe is mostly ejected at high velocity 
in a jet along the polar direction, while nearer the equatorial plane 
burning is less effective, and low-velocity O is mostly ejected.
We show that the unusual ratio of the width of the O and Fe nebular 
lines in SN~1998bw can be explained with a strongly aspherical 
explosion model viewed from a near-jet direction.  
Also, in this case the O line has a very sharp peak, in agreement with 
the observations. 

Much of our discussion was based on the identification of the 5200\AA\ feature
as a blend of [\FeII] lines. Although several caveats apply to that
identification, because other lines may contribute, we claim that [\FeII] lines
dominate. Other possible contributions are as follows. According to Sollerman
et al. (2000) the feature contains not only [\FeII] lines but also lines of
[\MgI], [\OI] and possibly [\FeI].  Lines of [\MgI] are included in our nebular
code, and the strength of the 5172\AA\ emission, which is consistent with the
5470\AA\ line reproducing the observed [\MgI] peak near 4500\AA\ is much
smaller than that of the [\FeII] lines. [\OI] 5577\AA\ is also included in our
code, and it is strong and fills up the red part of the emission, but it does
not contribute to the blue side (Mazzali et al. 2001).  As for 
[\FeI], Sollerman et al. (2000) say that their models `have too low a degree 
of ionization', suggest that the density in the Fe-emitting regions should 
be lower, and `regard the FeI emission as dubious'. We confirm this result.

Another possibility is allowed \FeII\ emission, such as identified 
by Filippenko (1989) in the Type II SNe 1987F and 1988I and by 
Filippenko \etal (1990) in the Type Ic SNe 1985F and 1987M. 
The main \FeII\ features are at 4570\AA\ (multiplets 37 and 38), 5190\AA\ 
(multiplet 42) and 5320\AA\ (multiplets 42 and 43). These lines are not 
included in our atomic model, because not all the collision strengths 
for these transitions are available. 
In particular multiplet 42 is not available. 
However, several arguments apply against the \FeII\ identification in SN~1998bw. 
1) In SNe 1987F and 1988I all three features are strong, while in the 
SNe 1985F and 1987M the feature at 4570\AA\ is narrow and it is identified 
as [\MgI], suggesting that the feature at 5200\AA\ could be [\FeII]. 
A similar situation holds for SN~1998bw, also a SN~Ic. 
2) \FeII\ emission occurs at high density. Filippenko (1989) infers
$\log n_e \sim 9$ -- 10 for SN~1987F. However, the densities we derive for 
SN~1998bw are more than one order of magnitude smaller (Mazzali \etal 2001). 
3) The relatively low density is confirmed by the large ratio ($\sim 4$) 
of the \CaII\ IR emission compared to [\CaII] 7300\AA. Fig.2 of Ferland
\& Persson (1989) suggests that $\log n_e < 9$. 
Finally, the models of Mazzali \etal (2001), which are based on [\FeII] only, 
reproduce the 5200\AA\ feature very well.  In any case, extending the model to 
include \FeII\ lines is a worthwhile effort which we are going to make. 

The 5200\AA\ feature is bluer on Nov. 26 and thereafter than on earlier epochs. 
Two principal factors are probably responsible for that: 1) the disappearance of 
the continuum, which is still significant in the earlier spectra; 2) the reduced 
intensity of \OI] 5577\AA, as shown also in our models (Mazzali et al. 2001). 
This is a high-density line, and it decreases quickly in strength as the ejecta 
expand. Fading \FeII \ emission may also play a role in causing a small 
wavelength shift. 

Weak unaccounted emission is present to the blue of the [\FeII] feature 
(4800-5100\AA). A similar emission is present in the nebular spectra 
of SNe~Ia as well (Patat \etal 2001), and it is poorly reproduced by 
synthetic spectra. It may be due to a forest of weak [\FeII] and [\CoII] 
transitions whose atomic parameters are not well known. 

The [\OI] line in other SNe~Ib/c (Filippenko \etal 1995, Matheson et 
al. 2001) also shows a strongly peaked profile, as in SN~1998bw. 
This probably signals the existence of oxygen at low velocity in most SNe Ib/c. 
Although there are very few SNe Ib/c in which the feature centered at 
5200\AA\ is detected with enough S/N, it appears that in these 
objects the [\OI] line has a similar profile. 
For low energy supernovae ($1 \times 10^{51}$ergs), the spherical model G 
shows a sharply peaked [\OI] line profile (Fig. 7). 
This seems to favor spherical explosions for these low energy supernovae. 
However, it is premature to conclude that asymmetry is completely absent in 
low energy SNIb/c, because the effects of asphericity in a late time spectra 
is not so large in case of normal supernovae as seen in Figure 7. 
This is due to the low expansion velocities. 
The oxygen line can be narrow even in the case of an aspherical 
explosion viewed from near the equator. 
For example, it is difficult to distinguish model D viewed from 75 deg 
from model G.
Because the probability of viewing an asphericl SN from the near polar 
direction is smaller than it is to view it near the equator, the observations
are not inconsistent with the possibility that most SNe Ib/c are 
more or less aspherical. 
More nebular spectra of SNe Ib/c with higher S/N ratio 
are needed for further investigation. 

Our aspherical explosion models may be able to explain the slow decline of 
the late light curve of SN 1998bw. 
In these models, the equatorial region is denser than in a 
spherically symmetric model with the same explosion energy. 
At advanced epochs this region may be able to trap $\gamma$-rays more
efficiently than a spherical model, as first suggested by Nakamura et
al. (2001a). 
Chugai (2000) showed that a spherical model could reproduce the light curve, 
if the density near the center (i.e. in the Fe-dominated region) was 
increased above that of the hydrodynamical model of Iwamoto \etal (1998). 
However, Sollerman et al (2000) find that the O-dominated region 
should be dense, and the Fe-dominated region not dense. 
Although taken individually these conclusions are probably correct,
they appear to be in conflict with one another. 
An aspherical model offers a natural solution, because it predicts the 
presence of high density, O-dominated matter near the center (Fig. 3). 

A small degree of linear optical polarization ($\sim 0.5$\%) was reported 
in SN 1998bw (Kay \etal 1998; Iwamoto \etal 1998; Patat \etal 2001). 
This can be explained with different combinations of asphericity and viewing 
angle, or with large scale clumping in a basically spherical envelope. 
One solution is that ejecta with a moderate departure from sphericity are 
viewed from slightly off the axis of symmetry. 
Our strongly aspherical explosion model has an axis ratio of about 3:2 at 
the outer edge of the oxygen envelope (Fig. 3), therefore it is consistent 
with the observed polarization, if it is viewed from near the jet direction. 

\bigskip

This work has been supported in part by the grant-in-Aid for COE
Scientific Research (07CE2002, 12640233) of the Ministry of Education,
Science, Culture, and Sports in Japan.

\clearpage

\begin{table}
\begin{center}
\begin{tabular}{cccccccccc}\hline\hline
$^{12}$C & 1.35E-01 & $^{13}$C & 1.84E-08 & $^{14}$N & 8.13E-05 & 
$^{15}$N & 2.63E-08 & $^{16}$O & 8.72\\
$^{17}$O & 1.11E-07 & $^{18}$O & 1.77E-06 & $^{19}$F & 1.45E-09 &
$^{20}$Ne& 5.38E-01 & $^{21}$Ne& 1.92E-03\\
$^{22}$Ne& 5.51E-02 & $^{23}$Na& 1.87E-02 & $^{24}$Mg& 3.35E-01 &
$^{25}$Mg& 4.08E-02 & $^{26}$Mg& 8.86E-02\\
$^{26}$Al& 6.28E-05 & $^{27}$Al& 7.06E-02 & $^{28}$Si& 5.27E-01 &
$^{29}$Si& 5.43E-02 & $^{30}$Si& 5.65E-02\\
$^{31}$P & 7.09E-03 & $^{32}$S & 2.37E-01 & $^{33}$S & 9.40E-04 &
$^{34}$S & 1.45E-02 & $^{36}$S & 1.71E-05\\
$^{35}$Cl& 3.92E-04 & $^{37}$Cl& 8.90E-05 & $^{36}$Ar& 4.01E-02 &
$^{38}$Ar& 5.73E-03 & $^{40}$Ar& 1.98E-07\\
$^{39}$K & 3.31E-04 & $^{40}$K & 7.56E-08 & $^{41}$K & 2.30E-05 &
$^{40}$Ca& 3.57E-02 & $^{42}$Ca& 1.67E-04\\
$^{43}$Ca& 1.25E-05 & $^{44}$Ca& 1.62E-03 & $^{46}$Ca& 1.23E-09 &
$^{45}$Sc& 2.39E-06 & $^{46}$Ti& 6.93E-05\\
$^{47}$Ti& 5.57E-05 & $^{48}$Ti& 2.02E-03 & $^{49}$Ti& 3.90E-05 &
$^{50}$Ti& 1.36E-09 & $^{50}$V & 4.60E-09\\
$^{51}$V & 9.91E-05 & $^{50}$Cr& 4.39E-04 & $^{52}$Cr& 8.25E-03 &
$^{53}$Cr& 7.72E-04 & $^{54}$Cr& 1.52E-07\\
$^{55}$Mn& 3.00E-03 & $^{54}$Fe& 3.87E-02 & $^{56}$Fe& 3.95E-01 &
$^{57}$Fe& 1.65E-02 & $^{58}$Fe& 4.88E-08\\
$^{59}$Co& 6.56E-04 & $^{58}$Ni& 2.71E-02 & $^{60}$Ni& 1.20E-02 &
$^{61}$Ni& 6.09E-04 & $^{62}$Ni& 5.08E-03\\
$^{64}$Ni& 4.29E-12 & $^{63}$Cu& 1.80E-05 & $^{65}$Cu& 1.12E-05 &
$^{64}$Zn& 1.60E-04 & $^{66}$Zn& 8.93E-05\\
$^{67}$Zn& 6.53E-07 & $^{68}$Zn& 1.23E-07 & $^{69}$Ga& 9.72E-10 &
$^{71}$Ga& 2.45E-10 & $^{70}$Ge& 1.02E-09\\\hline
\end{tabular}
\caption{Detailed Yields of Model C ($M_\odot$)
\label{tab:tab1}}
\end{center}
\end{table}

\begin{table}
\begin{center}
\begin{tabular}{cccccc}\hline\hline
{\bf Species}  &{\bf [X/O]\tablenotemark{a}}&{\bf [X/Fe]}  &{\bf  Species }
 & {\bf [X/O]}   &{\bf  [X/Fe]}\\\hline
$^{12}$C & -1.31  & -0.880  & $^{40}$Ca& -0.183  &  0.246\\
$^{14}$N & -4.09  & -3.66   & $^{44}$Ca&  0.0989 &  0.529\\
$^{16}$O &  0.000 &  0.430  & $^{45}$Sc& -1.17   & -0.740\\
$^{19}$F & -5.40  & -4.97   & $^{48}$Ti&  0.0153 &  0.445\\
$^{20}$Ne& -0.437 & -0.00731& $^{51}$V & -0.538  & -0.109\\
$^{23}$Na& -0.210 &  0.220  & $^{52}$Cr& -0.215  &  0.215\\
$^{24}$Mg& -0.146 &  0.284  & $^{55}$Mn& -0.605  & -0.176\\
$^{27}$Al&  0.127 &  0.557  & $^{54}$Fe& -0.224  &  0.206\\
$^{28}$Si& -0.0516&  0.378  & $^{56}$Fe& -0.430  &  0.000\\
$^{31}$P & -0.0192&  0.411  & $^{59}$Co& -0.668  & -0.238\\
$^{32}$S & -0.182 &  0.248  & $^{58}$Ni& -0.220  &  0.210\\
$^{35}$Cl& -0.769 & -0.339  & $^{63}$Cu& -1.46   & -1.03 \\
$^{36}$Ar& -0.244 &  0.186  & $^{64}$Zn& -0.750  & -0.320\\
$^{39}$K & -0.979 & -0.549  &          &         &       \\\hline\hline
\end{tabular}
\tablenotetext{a}{$\rm{[A/B]} \equiv \log_{10}(\rm{A/B}) 
- \log_{10}(\rm{A/B})_\odot$}
\caption{Abudances of major stable isotopes relative to 
the solar values for Model C
\label{tab:tab2}}
\end{center}
\end{table}

\clearpage

\begin{table}
\begin{center}
\begin{tabular}{cccccccc}\hline\hline
{\bf Model} & {\bf $\alpha/\beta$} & {\bf $E / 10^{51}$ergs }& 
{\bf $E_{\rm K}$ fraction\tablenotemark{a}} & 
{\bf angle\tablenotemark{b} [deg] } &
{\bf Fe FWHM\tablenotemark{c} [\AA]} &{\bf  O FWHM [\AA]} &
{\bf outer velocity [km s$^{-1}$]}\\\hline
 A & 16 & 10 & 0.65 &  5 & 376 & 150 & 9000\\
   &    &    &      & 15 & 344 & 150 & 8500\\
   &    &    &      & 30 & 304 & 150 & 8000\\
   &    &    &      & 45 & 248 & 150 & 6500\\
   &    &    &      & 60 & 216 & 150 & 5000\\
   &    &    &      & 75 & 144 & 150 & 4500\\
   &    &    &      & 85 & 144 & 150 & 4500\\\hline
 B &  8 & 30 & 0.5  &  5 & 352 & 150 & 10000\\
   &    &    &      & 15 & 344 & 150 & 9500\\
   &    &    &      & $> 30$ & & Too broad [\OI] & \\\hline
 C &  8 & 10 & 0.5  &  5 & 296 & 150 & 10000\\
   &    &    &      & 15 & 304 & 150 & 9500\\
   &    &    &      & 30 & 280 & 150 & 8500\\
   &    &    &      & 45 & 232 & 150 & 6500\\
   &    &    &      & 60 & 216 & 150 & 5000\\
   &    &    &      & 75 & 176 & 150 & 4500\\
   &    &    &      & 85 & 160 & 150 & 4000\\\hline
 D &  8 &  1 & 0.5  &  5 & 224 &  80 & 5000\\
   &    &    &      & 15 & 216 &  80 & 5000\\
   &    &    &      & 30 & 216 &  72 & 5000\\
   &    &    &      & 45 & 200 &  80 & 5000\\
   &    &    &      & 60 & 192 &  80 & 5000\\
   &    &    &      & 75 & 160 &  88 & 5000\\
   &    &    &      & 85 & 128 &  88 & 5000\\\hline
 E &  2 & 10 & 0.5  &  5 & 288 & 150 & 8500\\
   &    &    &      & 15 & 296 & 150 & 8000\\
   &    &    &      & 30 & 264 & 150 & 7000\\
   &    &    &      & $> 45$ & & Too broad [\OI] & \\\hline
 F &  1 & 10 & 0.5  &    &     & Too broad [\OI] & \\\hline
 G &  1 &  1 & 0.5  &    & 144 &  96 & 3500\\\hline\hline
\end{tabular}
\tablenotetext{a}{The ratio of the kinetic energy to $E$ 
initially deposited.}
\tablenotetext{b}{Angle between the line of sight and the jet direction.}
\tablenotetext{c}{Half line widths of the [\FeII]-blend (near 5200\AA) 
for SN 1998bw. The observed value is 380 \AA.}
\caption{Half Line Widths
\label{tab:tab3}}
\end{center}
\end{table}

\clearpage

\begin{figure}
\begin{center}
\epsscale{.7}
\plotone{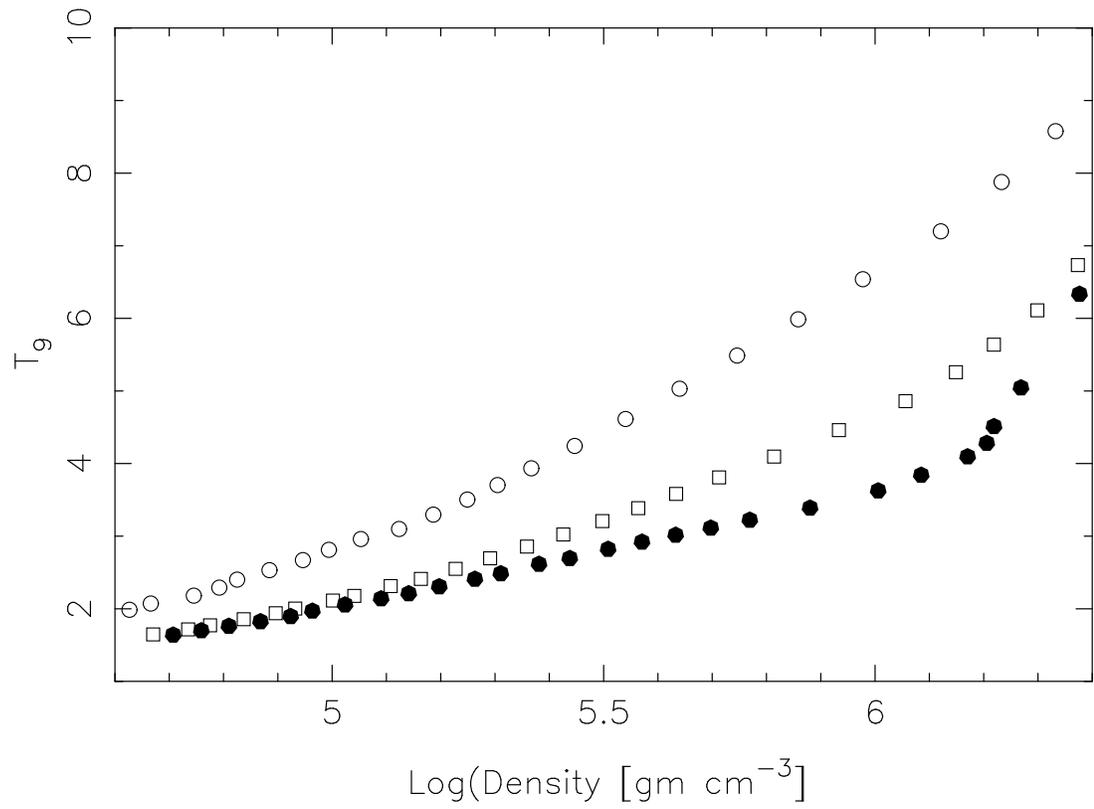}
\caption{The $\rho - T$ conditions of individual test particles at their 
temperature maximum (where $T_9 \equiv T \rm(K) / 10^9$). 
Fpr model C, the open circles denote those along the $z$-axis, and the filled 
circles denote those along the $r$-axis. The open squares denote those 
of model F (spherical model). 
\label{}}
\end{center}
\end{figure}

\begin{figure}
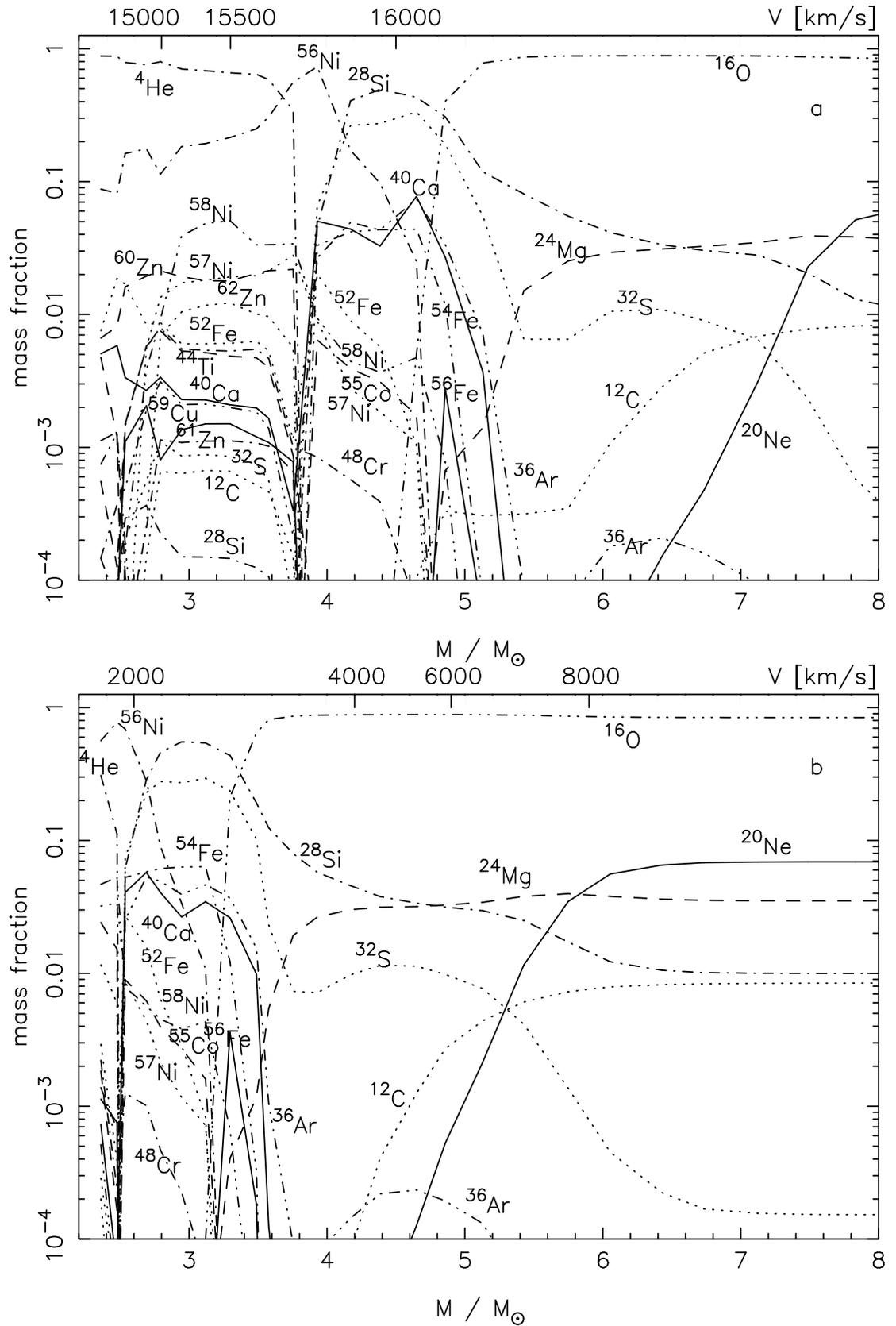

\begin{center}
\epsscale{.7}
\plotone{fig2a.epsi}\\
\plotone{fig2b.epsi}
\caption{The isotopic composition of the ejecta of model C in the direction of 
the jet (upper panel) and perpendicular to the jet (lower panel).
The ordinate indicates the initial spherical Lagrangian coordinate ($M_r$) 
of the test particles (lower scale), and 
the final expansion velocities ($V$) of those particles (upper scale).  
\label{fig:nuc1d}}
\end{center}
\end{figure}

\begin{figure}
\begin{center}
\epsscale{.7}
\plotone{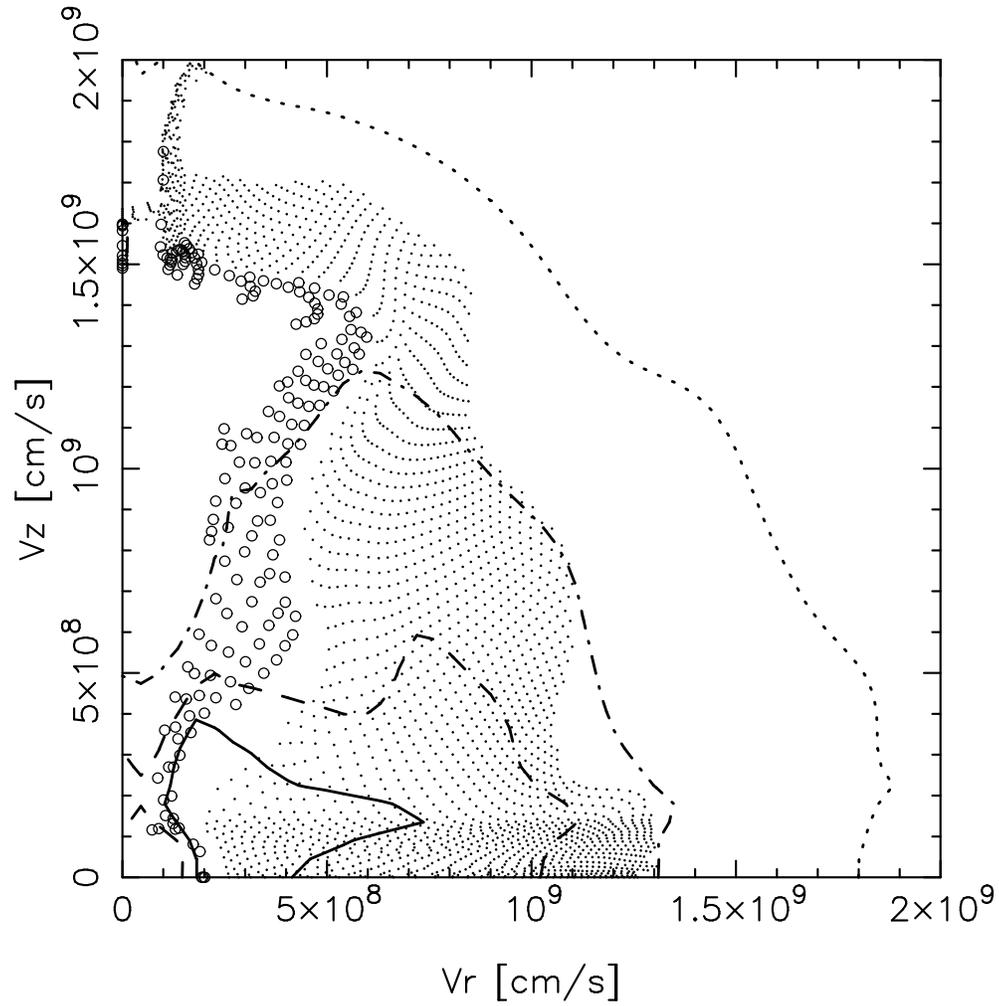}
\caption{The 2D distribution 
of $^{56}$Ni (open circles) and $^{16}$O (dots) of 
model C in the homologous expansion phase. 
Open circles and dots denote test particles in which the mass fraction 
of $^{56}$Ni and $^{16}$O, respectively, exceeds 0.1. 
The lines are density contours at the level of 0.5 (solid), 0.3 (dashed),
0.1 (dash-dotted), and 0.01 (dotted) of the max density, respectively. 
\label{fig:nuc2d}}
\end{center}
\end{figure}

\begin{figure}
\begin{center}
\epsscale{1.0}
\plotone{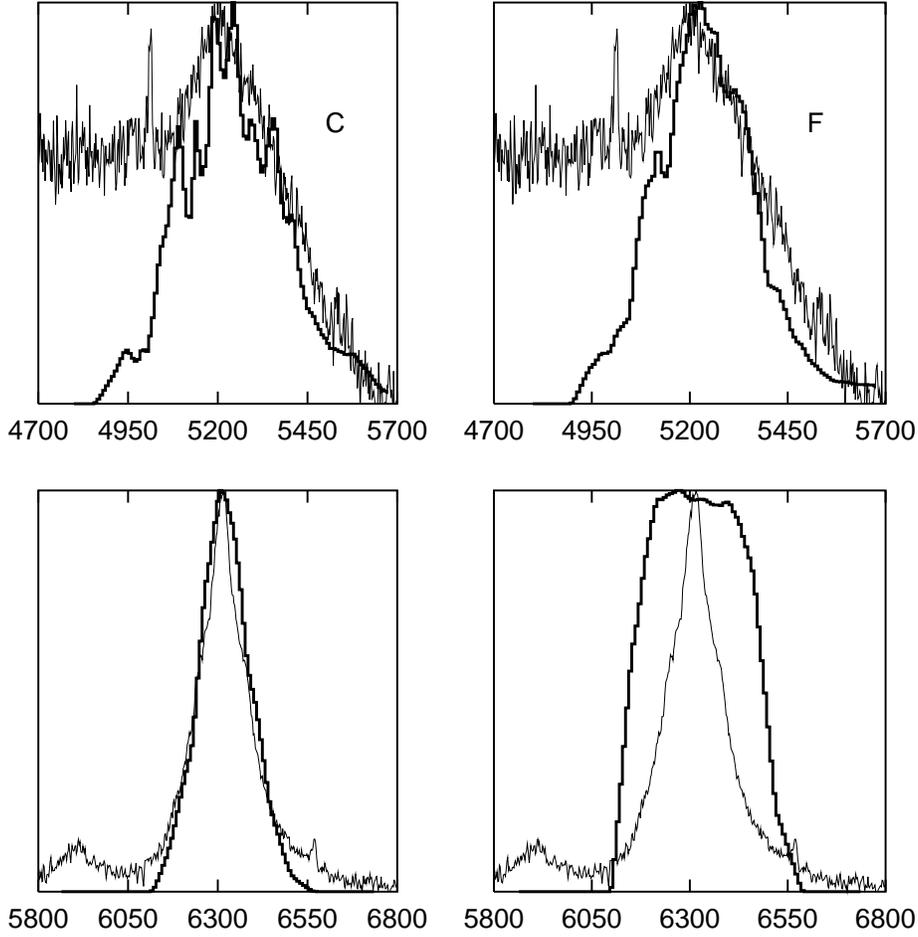}
\caption{The profiles of the [\FeII] feature (upper panels) and of  
[\OI] 6300, 6363 \AA\ (lower panels) for model C viewed 
at 15$^{\circ}$ from the jet direction (left panels; thick lines) and 
for model F (right panels). 
The observed lines at a SN rest-frame epoch of 216 days are also plotted for 
comparison (thin lines, Patat \etal 2001).
The intensities of the strongest lines, normalized to 
O {\small I}] 6300.3\AA\ are: 
[\FeII]\ 5158.8\AA: 0.122; 
[\FeII]\ 5220.1\AA: 0.026; 
[\FeII]\ 5261.6\AA: 0.083; 
[\FeII]\ 5273.3\AA: 0.039; 
[\FeII]\ 5333.6\AA: 0.060; 
[\FeIII]\ 5270.4\AA: 0.032; 
[\OI]\ 5577.3\AA: 0.022; 
and 
[\OI] 6363.8\AA: 0.330.  
\label{fig:line}}
\end{center}
\end{figure}

\begin{figure}
\begin{center}
\epsscale{1.0}
\plotone{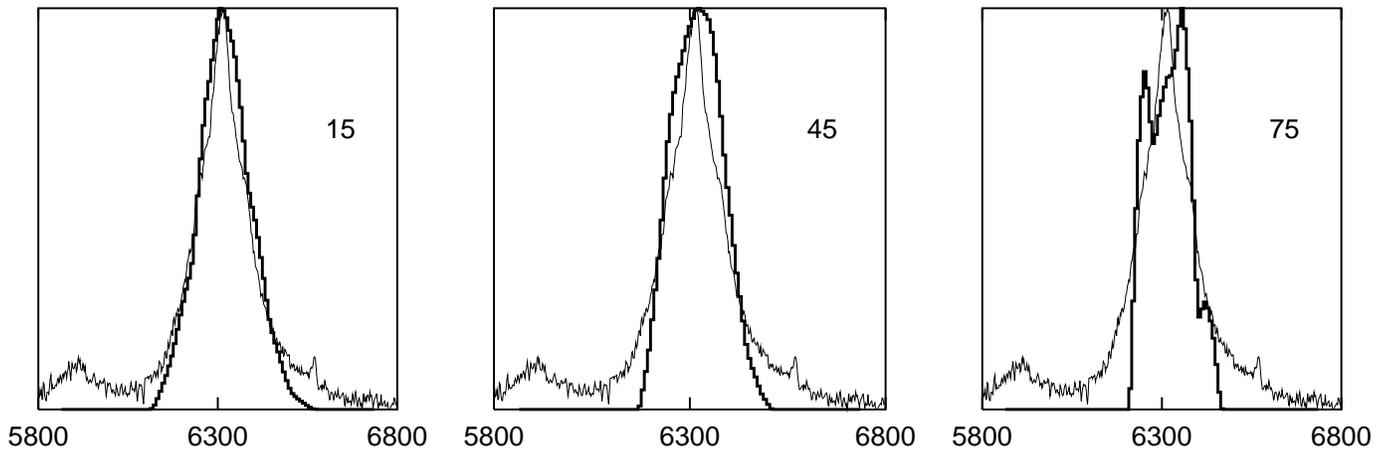}
\caption{The profiles of [\OI] 6300, 6363 \AA\ for model C with different 
orientation. The angles between the observer and the $z$-axis are 
15$^{\circ}$ (left), 45$^{\circ}$ (center) and 75$^{\circ}$ (right), 
respectively. 
\label{fig:oline}}
\end{center}
\end{figure}

\begin{figure}
\begin{center}
\epsscale{1.0}
\plotone{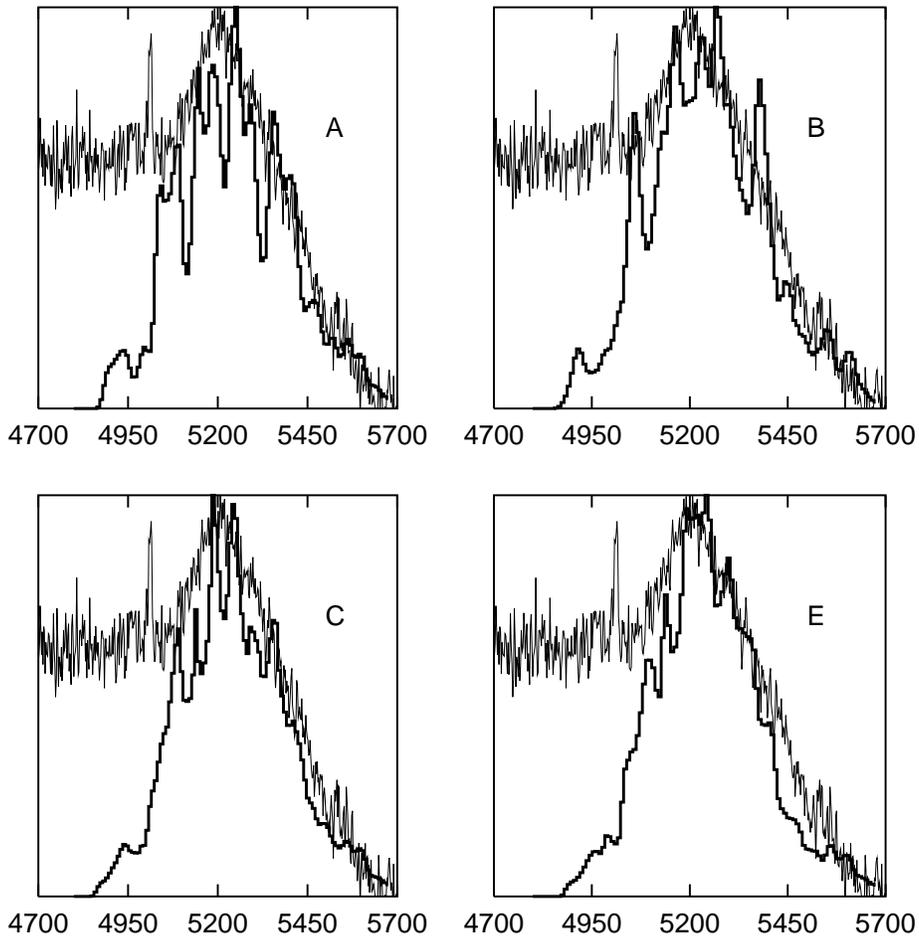}
\caption{The profiles of the [\FeII] feature for hypernove models viewed 
at 5$^{\circ}$. Each panel shows model A (upper-left), B (upper-right), 
C (lower-left), and E (lower-right), respectively. 
\label{fig:feline}}
\end{center}
\end{figure}

\begin{figure}
\begin{center}
\epsscale{1.0}
\plotone{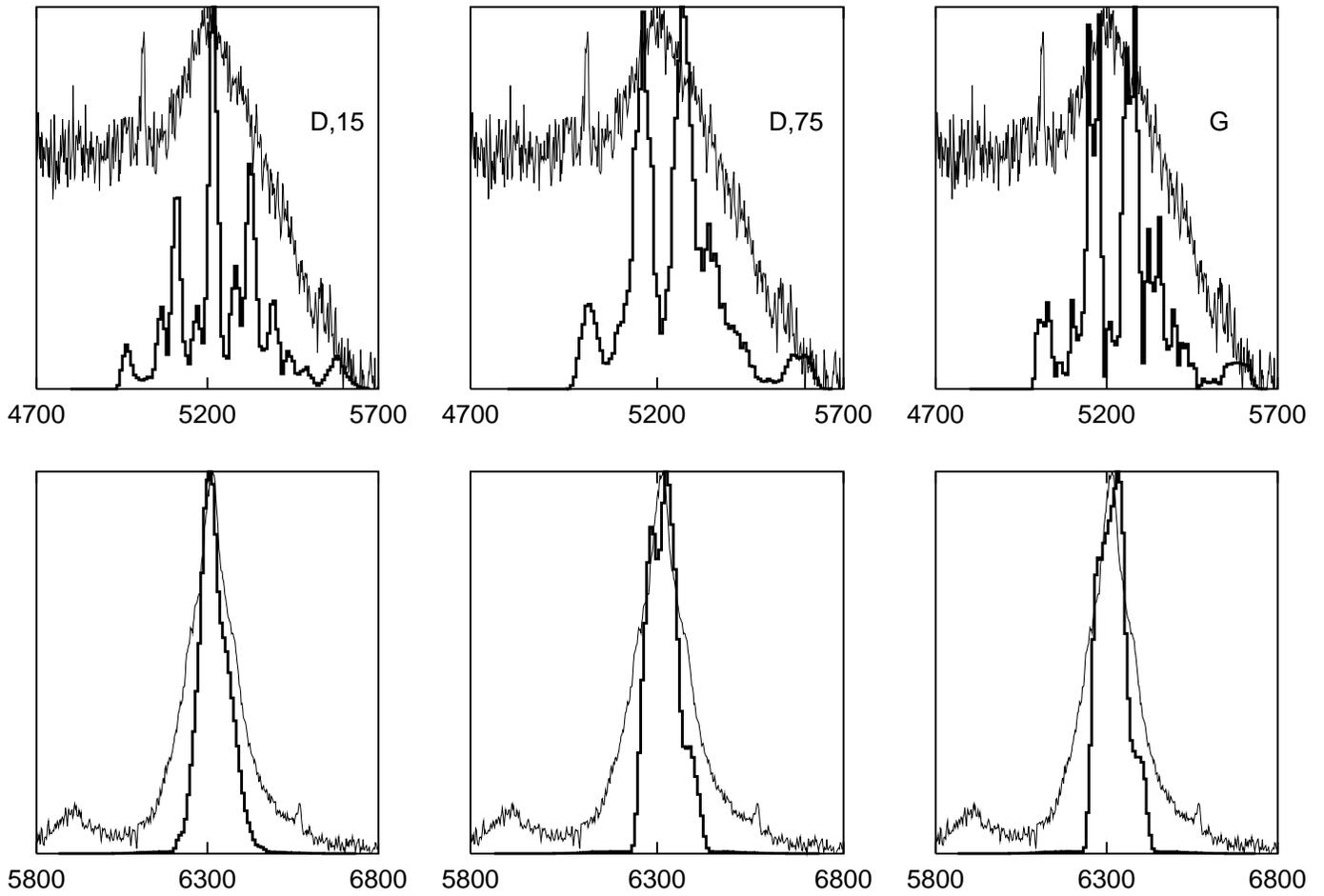}
\caption{The profiles of the [\FeII] feature (upper) and the [\OI] (lower) 
for normal energetic models. 
Those of model D viewed at 15$^{\circ}$ (left), 
75$^{\circ}$ (center), and model G are shown. 
\label{fig:1foeline}}
\end{center}
\end{figure}


\begin{thebibliography}{}

\bibitem{axelrod}Axelrod, T.S., 1980, Ph.D. Thesis, UCRL5294, 
Lawrence Livermore National Laboratory

\bibitem{branch}Branch, D. 2001, in ``Supernovae and Gamma Ray
Bursts,'' eds. M. Livio, \etal (Cambridge: Cambridge University Press),96

\bibitem{chugai} Chugai, N.N., 2000, Astronomy Letters, 26, 797 

\bibitem[Fer]{fer} Ferland, G.J., \& Persson, S.E., 1989, ApJ 347, 656 

\bibitem[Fil]{fil} Filippenko,~A.V., 1989, AJ 97, 726 

\bibitem{filippenko}Filippenko, A.V., Porter A.C., \& Sargent W.L.W., 
1990, AJ, 100, 1575 

\bibitem[F95]{f95} Filippenko,~A.V., et al., 1995, ApJ 450, L11 

\bibitem{galama}Galama, T.J., \etal 1998, Nature, 395, 670

\bibitem{hachisu92}Hachisu, I., Matsuda, T., Nomoto, K., \& Shigeyama, T. 
1992, ApJ, 390, 230

\bibitem{hachisu94}Hachisu, I., Matsuda, T., Nomoto, K., \& Shigeyama, T. 
1994, A\&AS, 104, 341

\bibitem{iwamoto98}Iwamoto, K., \etal 1998, Nature, 395, 672

\bibitem{kay}Kay, L.E., \etal, 1998, IAU Circ., No.6969

\bibitem{khokhlov}Khokhlov, A.M., H\"oflich, P.A., Oran, E.S., Wheeler, 
J.C., Wang, L., \& Chtchelkanova, A.Yu. 1999, ApJ, 524, L107

\bibitem{macfadyen}MacFadyen, A.I. \& Woosley, S.E. 1999, ApJ 524, 262

\bibitem[Mat01]{mat01} Matheson, T., Filippenko, A.V., Leonard, D.C.,
\& Shields, J.C., 2001, AJ 121, 1648 

\bibitem{mazzali}Mazzali, P.A., Nomoto, K, Patat, F., \& Maeda, K. 
2001, ApJ, 559, in press (astro-ph/0106095)

\bibitem{nagataki00}Nagataki, S. 2000, ApJS, 127,141

\bibitem{nakamura00a}Nakamura, T., Mazzali, P. A., Nomoto, K., \& Iwamoto, K. 
2001a, ApJ, 550, 991 

\bibitem{nakamura00b}Nakamura, T., Umeda, H., Iwamoto, K., Nomoto, K., 
Hashimoto, M., Hix, R.W., \& Thielemann, F.-K. 2001b, ApJ, 555, 880

\bibitem{nomoto88}Nomoto, K., \& Hashimoto, M. 1988, Phys. Rep., 256, 173

\bibitem{nomoto01a}Nomoto, K., \etal 2001a, in ``Supernovae and Gamma Ray
Bursts,'' eds. M. Livio, \etal (Cambridge: Cambridge University Press), 144 
(astro-ph/0003077)

\bibitem{nomoto01b}Nomoto, K., \etal 2001b, in `` The influence of Binaries 
on Stellar Population Studies,'' ed. D. Vanbeveren (Kluwer), 507,  
(astro-ph/0105127)

\bibitem{paczynski}Paczynski, B. 1998, ApJ, 494, L45

\bibitem{patat}Patat, F., \etal 2001, ApJ, 555, 900

\bibitem{sollerman}Sollerman, J., Kozma, C., Fransson, C., Leibundgut, B., 
Lundqvist, P., Ryde, F., \& Woudt, P. 2000, ApJ 537, L127

\bibitem{tf}Thielemann, F.-K., Nomoto, K., \& Hashimoto, M. 1996, ApJ, 460, 408

\bibitem{ws}Woosley, S.E., Eastman, R.G., \& Schmidt, B.P. 1999, ApJ, 516, 788

\bibitem{ws2}Woosley, S.E. 1993, ApJ, 405, 273

\end{thebibliography}
\end{document}